# Effect of Phonon Scattering on Intrinsic Delay and Cut-Off Frequency of Carbon Nanotube FETs

Youngki Yoon, Yijian Ouyang, and Jing Guo
Department of Electrical and Computer Engineering
University of Florida, Gainesville, FL, 32611, USA

## ABSTRACT

The effect of phonon scattering on the intrinsic delay and cut-off frequency of Schottky barrier CNTFETs is examined. Carriers are mostly scattered by optical and zone boundary phonons beyond the beginning of the channel. We show that the scattering has a small direct effect on the DC on-current of the CNTFET, but it results in significant decrease of intrinsic cut-off frequency and increase of intrinsic delay.

I.    Introduction

Excellent carrier transport properties of single wall carbon nanotubes (CNTs) lead to a strong interest in potential applications of CNTFETs in digital and radio-frequency (RF) electronics. Extensive experiments have been performed recently to explore AC characteristics of CNTFETs [1-3]. Theoretical calculations of the cut-off frequency and delay time of CNTFETs have been based on the assumption of ballistic transport, which predicts THz operation [4-7]. Little, however, is known about how scattering affects the speed and high frequency performance of CNTFETs. The understanding is essential for assessing the performance potential of CNTFETs for electronics applications.

The effect of scattering on the DC current of metallic and semiconducting CNTs has been previously examined [8]. The dominant scattering mechanism in a high-quality CNT is phonon scattering. At low biases, acoustic phonon (AP) scattering with a mean free path (mfp) ~1μm is dominant, and at high biases, optical phonon (OP) and zone boundary (ZB) phonon scattering with a mfp of ~10nm is most important. A CNTFET with a channel length several times longer than phonon scattering mfp can still deliver a near ballistic DC on-current [9]. In this work, the effect of phonon scattering on the intrinsic delay and cut-off frequency of CNTFETs is examined. We show that although its direct effect on on-current and transconductance is small, phonon scattering results in significant (i) pile-up of charge in the channel and increase of the intrinsic gate capacitance, (ii) random walks of carriers in the channel and decrease of the average carrier velocity, (iii) decrease of the intrinsic cut-off frequency and increase of the intrinsic delay. This study indicates the different role of phonon scattering on the DC on-current and on the intrinsic speed and high-frequency performance of CNTFETs. Even for a CNTFET that delivers a near ballistic DC on-current, the effect of phonon scattering on intrinsic delay and cut-off frequency can be significant.

**II.    Approach**

We simulated a coaxially gated Schottky barrier (SB) CNTFET at room temperature (T=300K). The coaxial gate geometry offers the best electrostatic gate control. The nominal device has a 5nm $HfO_2$ gate oxide with a dielectric constant of 16. The diameter of the (17, 0) CNT channel is $d \approx 1.33\ nm$, which results in a bandgap of $E_g \approx 0.63\ eV$. A power supply voltage of 0.5V is assumed, which is close to the value specified for the end-of-roadmap FETs in the ITRS roadmap. A nominal channel length of 100nm is used, and it is varied to explore the channel length effect. The metal source/drain is directly attached to the CNT channel, and the Schottky barrier height between the source/drain and the intrinsic CNT channel is $\Phi_{Bn} = 0$.

The DC characteristics of CNTFETs are simulated by solving the Schrödinger equation using the non-equilibrium Green's function (NEGF) formalism self-consistently with the Poisson equation. We treat scattering by different phonon modes in the NEGF simulation with the self-consistent Born approximation [9, 10] using the perturbation potential computed by Mahan [11]. An atomistic description of the nanotube using a tight binding Hamiltonian with a p$_z$ orbital basis set is applied. The atomistic treatment is computationally expensive in real space, but significant saving of computational cost can be achieved by the mode space approach.

We compute the intrinsic delay and cut-off frequency of the device at the ballistic limit and in the presence of phonon scattering (assuming zero parasitic capacitance). The design of CNTFETs in experiments to date has not been optimized for high-frequency performance, which is limited by parasitic capacitances between electrodes. The recent progress on quasi one-dimensional nanowire contacts and on the CNT array channel, however, may significantly reduce the parasitic capacitance and eventually lead to a performance close to the intrinsic limit. Nevertheless, this work focuses on how phonon scattering affects the intrinsic performance of CNTFETs, which sets the ultimate limit of CNTFETs for digital and RF electronics applications.

The cut-off frequency (the unity current gain frequency), $f_T$, is an important performance metric for high frequency performance of a transistor. We compute the intrinsic $f_T$ using the quasi-static approximation [2, 4-7],

$$f_T = \frac{1}{2\pi} \frac{g_m}{C_g}\bigg|_{V_D=V_{DD}} = \frac{1}{2\pi} \frac{g_m \cdot \partial V_G}{C_g \cdot \partial V_G}\bigg|_{V_D=V_{DD}} \approx \frac{1}{2\pi} \frac{\partial I_D}{\partial Q_{ch}}\bigg|_{V_D=V_{DD}} \quad (1)$$

where $g_m$ is the transconductance, $C_g$ is the intrinsic gate capacitance, $I_D$ is the source-drain current, $Q_{ch} = q\int_0^{L_{ch}} N_e(x)dx$ is the total charge in the CNT channel, $N_e(x)$ is the electron density as a function of the channel position, and $L_{ch}$ is the channel length. The derivative, $\partial I_D / \partial Q_{ch}$, in Eqn. (1) is obtained by running the detailed DC simulations and computing the ratio of the current variation to the channel charge variation at two slightly different gate voltages and $V_D$=0.5V.

The intrinsic delay, which characterizes how fast a transistor intrinsically switches, is an important performance metric for digital electronics applications. The intrinsic delay, $\tau = (Q_{on} - Q_{off})/I_{on}$, is computed by running DC simulations at the off-state ($V_G$=0, $V_D$=0.5V) and the on-state ($V_D$=$V_G$=0.5V), where $Q_{on}$ and $Q_{off}$ are the total charge in the channel at on-state and off-state, respectively, and $I_{on}$ is the on-current.

**III.　Results**

*3.1 On-current*

Fig. 1a shows the simulated $I_D$ vs. $V_G$ characteristics for $V_D$=0.5V at the ballistic limit (the dashed line) and in the presence of scattering (the solid line). In the presence of phonon scattering, the CNTFET delivers a near ballistic DC on-current (~85% of the ballistic value) and the transconductance, $g_m = \partial I_D / \partial V_G$ is also close to the ballistic value even though the channel length is several times longer than the mfp of OP and ZB phonon scattering. This can be explained as follows [9, 10]. An electron injected from the source can be accelerated by the electric field and obtain a kinetic energy larger than the OP/ZB phonon energy. It can then emit an OP/ZB phonon and get backscattered, but because the OP/ZB phonon energy in CNT is large (>160meV), the backscattered carrier faces a much thicker and higher barrier, and has little chance to tunnel back to the source. The result is that the electron rattles around in the channel and finally exits to the drain. OP and ZB phonon scattering has a small direct effect on the DC on-current of the CNTFET under modest gate biases.

Fig. 1b plots the on-current versus the channel length. In the presence of phonon scattering, the CNTFET delivers about 95% of the ballistic on-current at a channel length of 20nm and about 80% at 200nm. The ballistic on-current is nearly independent of the channel length, and the on-current in the presence of phonon scattering slightly decreases as the channel length increases, mostly due to the increasingly important role of near elastic acoustic and radial breathing mode (RBM) phonon scattering. OP and ZB scattering with a short mfp has a small direct effect on the on-current. The CNTFET delivers a DC on-current $\geq 80\%$ of the ballistic value for a channel length up to 200nm.

*3.2 charge and velocity*

Though its effect on DC on-current and transconductance is small, phonon scattering can result in significant decrease of the average carrier velocity, pile-up of charge in the channel, and increase of the intrinsic gate capacitance. Fig. 2a plots the average electron velocity at on-state as a function of the channel position, which is computed as $\upsilon(x) = I_{on}/[qN_e(x)]$, where $N_e(x)$ is the electron density at on-state. At the ballistic limit, the electron velocity increases at the beginning of the channel and then remains nearly constant at about $6 \times 10^7 cm/s$. Because the first subband bends at the beginning of the channel and then remains nearly flat, electrons only populate the +$k$ states with high kinetic energies beyond the beginning of the channel, which have a large band-structure-limited velocity. Phonon scattering, however, significantly reduces the electron velocity in the channel due to the following two reasons. First, phonon scattering backscatters electrons and results in population of the –k states. Second, OP and ZB phonon scattering results in a decrease of electron kinetic energy. A large percent of the current is delivered by the electrons at the bottom of the subband, where the velocity of electrons is much smaller.

Fig. 2b plots the electron density $N_e(x)$ at on-state versus the channel position. The electron density at the ballistic limit remains nearly constant in the channel because the subband profile is nearly flat due to good gate electrostatic control. It reaches peak values at the metal/CNT contacts due to metal induced gap states (MIGS) at the ends of the channel. Phonon scattering results in a significant increase of the electron density in the channel because carriers are scattered by phonons and rattle around in the channel region.

*3.3 Intrinsic cut-off frequency and delay*

Fig. 3a plots the intrinsic cut-off frequency versus the channel length at the ballistic limit (the circles) and in the presence of phonon scattering (the crosses). The dashed line is a fitting of the ballistic result by $f_T = 110 GHz \cdot \mu m / L_{ch}$, and the solid line is a fitting of the scattering result by $f_T = 40 GHz \cdot \mu m / L_{ch}$. Although the transistor delivers a near ballistic DC on-current (>80%), the cut-off frequency in the presence of phonon scattering is only about 40% of the ballistic value for a channel length of $L_{ch}$=50-200nm. The reason is that random walks of electrons due to phonon scattering result in pile-up of charge in the channel and significant increase of the intrinsic gate capacitance, although its effect on the DC current and transconductance is small. As channel length decreases from 50nm to 20nm[12], the ratio of $f_T$ in the presence of phonon scattering to that at the ballistic limit increases to about 54% due to reduced OP/ZB scattering and quasi-ballistic transport. In a previous work [5], we showed that the $f_T$ of a ballistic CNTFET is about 1.5 times larger than that of a ballistic Si MOSFET at the same channel length due to a larger band-structure-limited velocity. Because a nanoscale Si MOSFET operates at 40-50% of its ballistic limit [13], we expect scattering decreases the intrinsic $f_T$ of a Si MOSFET by a factor of 0.4-0.5.

Fig. 3b plots the intrinsic delay of the CNTFET vs. the channel. The dashed line is a linear fitting to the ballistic result by $\tau = L_{ch} \times 1.71 ps/\mu m$. The intrinsic delay at the ballistic limit scales linearly with the channel length for $L_{ch}$>20nm. In the presence of phonon scattering, the intrinsic delay significantly increases because phonon scattering lowers the average carrier velocity in the channel. The intrinsic delay in the presence of phonon scattering at $L_{ch}$=200nm is about 210% larger than that at the ballistic limit. When the channel length decreases to a value close the phonon scattering mfp, the speed degradation by scattering is less severe. At a channel length of 20nm, the intrinsic delay with scattering is only about 40% larger than that at the ballistic limit.

## IV. Discussions and Conclusions

The modeled CNTFET has a thin coaxial high-κ gate insulator and operates close to the quantum capacitance limit [14]. We show that the direct effect of phonon scattering on the DC source-drain current and transconductance is small. The indirect effect of phonon scattering through self-consistent electrostatics is also small because of the good gate electrostatic control [9]. The transistor delivers a near ballistic on-current. On the other hand, the intrinsic gate capacitance $C_g$, which is the series combination of the gate insulator capacitance $C_{ins}$ and the CNT quantum (semiconductor) capacitance $C_Q$, is close to $C_Q$ at the quantum capacitance limit ($C_{ins} \gg C_Q$) [14]. Phonon scattering results in a significant increase of $C_Q$, which leads to an increase of $C_g$ and decrease of the intrinsic cut-off frequency, $f_T = g_m/(2\pi C_g)$.

We also examined the case when the gate insulator is thick and the transistor operates close to the conventional MOSFET limit ($C_{ins} \ll C_Q$ so that $C_g \approx C_{ins}$) [14]. In contrast, phonon scattering does not lead to an increase of the intrinsic gate capacitance at the MOSFET limit. It, however, can significantly lower the DC source-drain current and transconductance through an indirect self-consistent electrostatic effect [9]. The intrinsic cut-off frequency still decreases, but due to the decrease of the DC transconductance, instead of the increase of the intrinsic gate capacitance.

**Acknowledgment**

It is our pleasure to thank S. Hasan and Prof. M. Lundstrom of Purdue University for discussions. The computational facility was supported from the NSF Grant No. EIA-0224442, IBM SUR grants and gifts, and a DURIP Grant from ARO.

**FIGURES**

Fig. 1. (a) $I_D$ vs. $V_G$ at $V_D$=0.5V at the ballistic limit (the dashed line) and in the presence of phonon scattering (the solid line). The channel length is 100nm. (b) The on-current (at $V_D$=$V_G$=0.5V) versus the channel length at the ballistic limit (the dashed line) and in the presence of phonon scattering (the solid line). The inset of (b) sketches the first subband profile at on-state. Electrons injected from the source can emit an OP or ZB phonon beyond the beginning of the channel, but such scattering has a small direct effect on the source-drain current.

Fig. 2. (a) The average carrier velocity and (b) the electron density versus the channel position at $V_D$=$V_G$=0.5V at the ballistic limit (the dashed lines) and in the presence of phonon scattering (the solid lines). The large electron density at the ends of the channel is due to MIGS.

Fig. 3. (a) The cut-off frequency versus the channel length at on-state ($V_D$=$V_G$=0.5V). The circles are numerically computed $f_T$ at the ballistic limit and the dashed line is a fitting curve of $f_T = 110 GHz \cdot \mu m / L_{ch}$. The crosses are numerically computed $f_T$ in the presence of phonon scattering and the solid line is a fitting curve of $f_T = 40 GHz \cdot \mu m / L_{ch}$. (b) The intrinsic delay versus the channel length at the ballistic limit (the circles) and in the presence of phonon scattering (the crosses). The dashed line is a linear fitting of the ballistic result by $\tau = L_{ch} \times 1.71 ps / \mu m$.

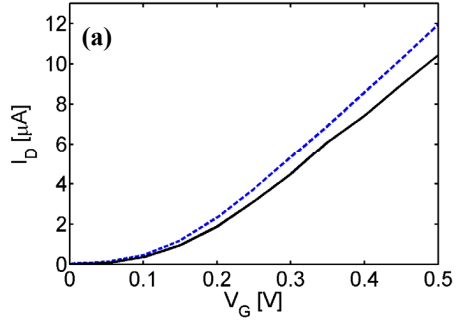 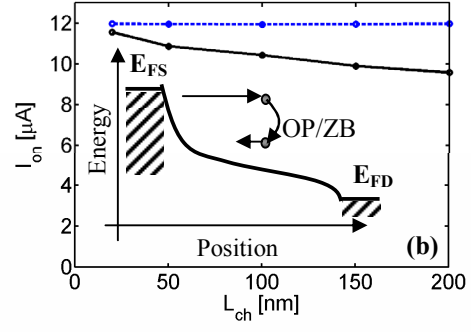

Fig. 1.

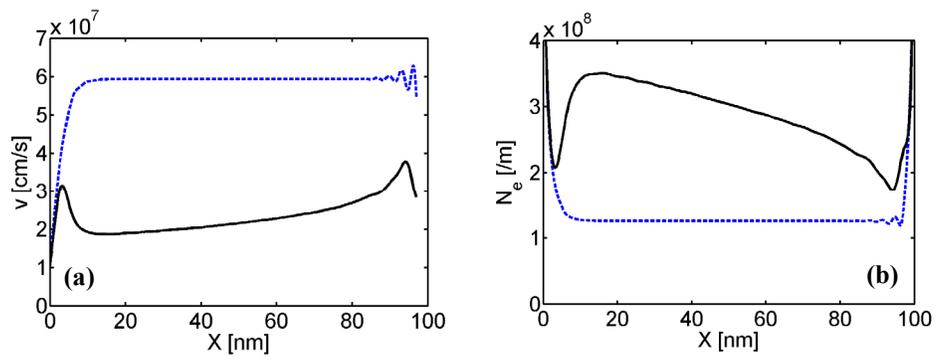

Fig. 2.

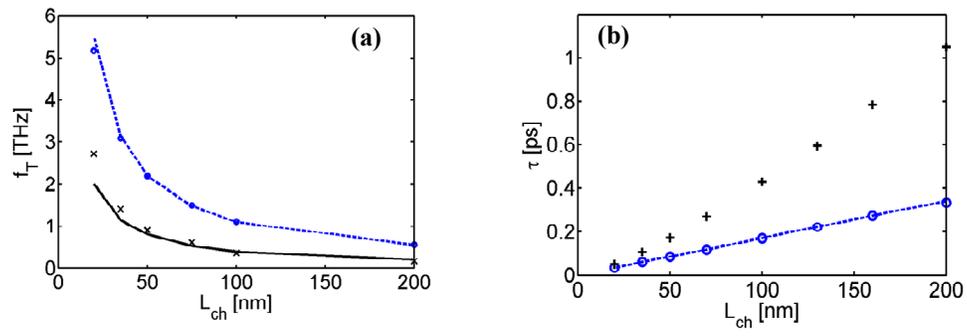

Fig. 3.